\documentclass[useAMS,usenatbib]{mn2e}
\usepackage{rotating}
\usepackage{journals}
\usepackage{times}

\def\approxgt{\ifmmode \rlap{$>$}{}_{{}_{{}_{\textstyle\sim}}} \else%
$\rlap{$>$}{}_{{}_{{}_{\textstyle\sim}}}$\fi} 
\def\approxlt{\ifmmode \rlap{$<$}{}_{{}_{{}_{\textstyle\sim}}} \else%
$\rlap{$<$}{}_{{}_{{}_{\textstyle\sim}}}$\fi}

\def\arcsec{\hbox{$^{\prime\prime}$}}

\def\src{1A~1246--588}
\def\ecs{erg~cm$^{-2}$s$^{-1}$}

\LARGE \normalsize \title[A kilohertz QPO in \src]{Detection of a 1258 Hz high--amplitude kilohertz
quasi--periodic oscillation in the ultra-compact X--ray binary \src}

\author[Jonker et al.]  {P.G.~Jonker$^{1,2,3}$\thanks{email :
p.jonker@sron.nl}, J.J.M. in 't Zand$^{1,3}$, M. M\'endez$^{1,3,4}$, M. van der Klis$^{4}$\\
$^1$SRON, Netherlands Institute for Space Research, Sorbonnelaan 2, 3584~CA, Utrecht, The Netherlands\\
$^2$Harvard--Smithsonian  Center for Astrophysics, 60 Garden Street, Cambridge, MA~02138, Massachusetts,
U.S.A.\\
$^3$Astronomical Institute, Utrecht University, P.O.Box 80000, 3508 TA, Utrecht, The Netherlands\\
$^4$Astronomical Institute ``Anton Pannekoek'', University of Amsterdam, Kruislaan 403, 1098 SJ Amsterdam\\
}

\begin{document}

\maketitle

\begin{abstract} \noindent  We have observed the ultra-compact low--mass X--ray binary (LMXB) \src\ with the {\it
Rossi} X--ray Timing Explorer (RXTE). In this manuscript we report the discovery of a kilohertz quasi--periodic
oscillation (QPO) in \src. The kilohertz QPO was only detected when the source was in a soft high--flux state
reminiscent of the lower banana branch in atoll sources. Only one kilohertz QPO peak is detected at a relatively high
frequency of 1258$\pm$2 Hz and at a single trial significance of more than 7~$\sigma$.  Kilohertz QPOs with a higher
frequency have only been found on two occasions in 4U~0614+09. Furthermore, the frequency is higher than that found for
the lower kilohertz QPO in any source, strongly suggesting that the QPO is the upper of the kilohertz QPO pair often
found in LMXBs. The full--width at half maximum is 25$\pm$4 Hz, making the coherence the highest found for an upper
kilohertz QPO. From a distance estimate of $\approx$6 kpc from a radius expansion burst we derive that \src\ is at a
persistent flux of $\approx$0.2--0.3 per cent of the Eddington flux, hence \src\ is one of the weakest LMXBs for which
a kilohertz QPO has been detected. The root--mean--square (rms) amplitude in the 5--60 keV band is 27$\pm$3 per cent,
this is the highest for any kilohertz QPO source so far, in line with the general anti--correlation between source
luminosity and rms amplitude of the kilohertz QPO peak identified before. Using the X--ray spectral information we
produce a colour--colour diagram. The source behaviour in this diagram provides further evidence for the atoll nature
of the source.   

\end{abstract}


\section{Introduction} 
 
Ultra-compact X--ray binaries (UCXBs) are low--mass X--ray binaries (LMXBs) that have an orbital period shorter
than $P_{\rm orb}\approx$1~hr, implying such a small Roche lobe that the donor in an UCXB will have lost (most of)
its hydrogen (\citealt{1986ApJ...304..231N}; \citealt{1986AA...155...51S}). For 12 LMXBs $P_{\rm orb}$ has been
measured to be in the ultra-compact regime (\citealt{2006astro.ph..5722N}). 

In the power spectra of $\sim$20 accreting neutron star LMXBs kilohertz quasi--periodic oscillations
(QPOs) have been discovered (see \citealt{2006csxs.book...39V} for a review). Kilohertz QPOs are thought to be
caused by motion of matter a few kilometres above the surface of accreting neutron stars. Even though there is as
yet no agreement about the exact physical mechanism causing the X--ray light curves to be modulated at kilohertz
frequencies, most models agree that the frequency of one of the observed kilohertz QPOs reflects the frequency of
orbital motion at the inner edge of the accretion disc. Hence, the kilohertz QPOs potentially allow one to detect
effects of the strong gravitational fields and to constrain the neutron star mass--radius relation
(e.g.~\citealt{milaps1998}). 

Often two kilohertz QPO peaks separated by $\Delta\nu$=200--360\,Hz are found (again see \citealt{2006csxs.book...39V}
for a review). The highest frequencies have been observed in the UCXB 4U~0614+09 ($1329\pm4$~Hz, \citealt{vafova2000}
and 1273.6$\pm$9.5\,Hz; \citealt{2002ApJ...568..912V}). At high ($\approxgt$400 Hz) neutron star spin frequencies $\Delta
\nu$ has been found to be equal to half the spin frequency (e.g.~in SAX~J1808.4--3658; \citealt{2003Natur.424...44W}),
whereas in sources where the neutron star spin is $\approxlt$350 Hz $\Delta \nu$ is found to be close to the spin
frequency (see \citealt{2006AdSpR..38.2675V}, \citealt{2006csxs.book...39V} and \citealt{2006csxs.book..113S} for
reviews).

\src\ was discovered with the Ariel--V observatory in the mid 1970s at a level of roughly
$(1-2)\times10^{-10}$~\ecs\ (2--10 keV) by \citet{1976MNRAS.177P..13S}, but it had also been detected by UHURU a few
years earlier (\citealt{1978ApJS...38..357F}). The Ariel--V data exhibited a flare with a peak 5 times above the
lowest level (\citealt{1977MNRAS.179P..27C}). On Feb. 16, 1985, the first pointed observation of \src\ was taken
with EXOSAT. From the HEASARC standard products we found that the light curve is featureless and the 2--10 keV
flux is $9.5\times10^{-11}$~\ecs. In 1997, the first X--ray burst was detected with the BeppoSAX Wide Field
Cameras (WFCs; \citealt{1997IAUC.6538....2P}) from a position consistent with that of \src\
(\citealt{1997IAUC.6546....1B}). The increased localisation accuracy provided by the WFCs enabled an
identification with a ROSAT source and thus the position is known with an accuracy of 9\arcsec\ (Boller et al.
1997). This enabled \citet{2006A&A...446L..17B} to identify the optical counterpart and determine an UCXB nature
through the ratio of the X--ray to optical flux. In August 2006 a type--I X--ray burst was detected with the
Swift Burst Alert Telescope (\citealt{2006GCN..5436....1R}; \citealt{2006ATel..875....1K}) and the satellite
autonomously slewed to the target within 193~s. The burst was still detectable with the Swift X--ray telescope
and the total duration could be accurate determined to be 28 minutes. Such a duration is similar to the long
bursts from other UCXBs (see \citealt{2007astro.ph..1810I}).

Here, we report on our analysis of {\it Rossi} X--ray Timing Explorer (RXTE) observations of \src.

\section{Observations, analysis and results} 

\begin{figure}
  \includegraphics[angle=0,width=8cm,clip]{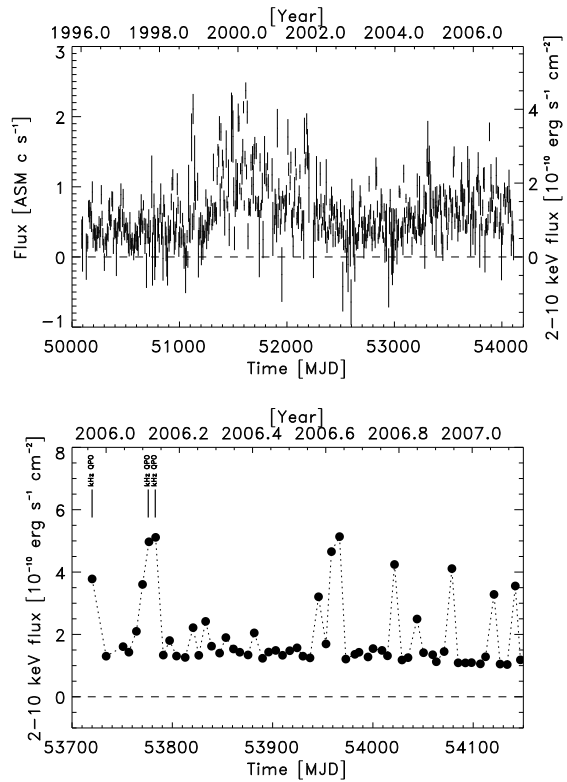}
  
  \caption{{\it Top panel:} The RXTE All Sky Monitor 7--day--average light curve of \src. The ordinate on the
  left/right hand side gives the source flux in ASM counts per second/2--10 keV in \ecs. The flare episodes where
  the count rate increases by a factor $\approx$3 can be discerned.  {\it Bottom panel:} The observed (absorbed)
  2--10 keV flux derived from the PCA monitoring observations of \src\ fitting an absorbed  blackbody plus
  power--law model to the Standard 2 data. We have also indicated during which observations the kilohertz QPO was
  significantly detected. Note that the abscissa of the top and bottom panel are different and that the size of
  the data points is larger than the error on the data points in the {\it bottom panel}.}

\label{rxtelc}
\end{figure}

We have used proportional counter array (PCA) data from 55 short RXTE observations of \src. All data obtained between MJD
53720--54108 (UTC; Dec 16, 2005--Jan.~15, 2007) amounting to a total of $\approx45$ ksec were taken from program P90042. In
the {\it top panel} of Figure~\ref{rxtelc} we have plotted the light curve obtained with the RXTE All Sky Monitor (ASM).
Besides ASM count rate we show the source flux in \ecs; for this we used the fact that the \src\ PCA X--ray spectrum in the
low--flux state resembles that of the Crab. The source is persistently detected. It remained at a level of roughly 0.37 ASM
counts s$^{-1}$ which is 5~mCrab (2--12 keV), except for a long period between Oct.~1998 (MJD~51100) and Oct.~2001
(MJD~52200) when it exhibited much more variability and the flux went up to 2.4 ASM counts~s$^{-1}$ (32 mCrab) and
similarly after mid 2004. There are flare--like features visible with peak fluxes that are $\approx$3 times higher than
normal, last a couple of weeks and have a recurrence time between $\approx$20 and 100 days. In the  {\it bottom panel} of
Figure~\ref{rxtelc} we show the absorbed 2--10 keV flux as derived from fitting an absorbed power--law plus blackbody model
to the X--ray spectra extracted from the RXTE/PCA Standard 2 data (using {\sl ftools 6.0.4} and {\sl xspec 12.2.1};
\citealt{ar1996}). 

Using 16~s--long segments of high time resolution PCA data (122$\mu$s resolution), we calculated power spectra up to a
Nyquist frequency of 4096~Hz in the full energy band of 2--60 keV. We first combined all the power spectra from all data.
The high--frequency (256--4096 Hz) part of the average power spectrum was searched for the presence of kilohertz QPOs. We
did not detect a kilohertz QPO. However, we next selected the data for which the total PCA count rate was above $\sim$150
counts s$^{-1}$. A strong kilohertz QPO was discovered. Subsequent subselections showed that the kilohertz QPO was
significantly present only in the flare data before MJD 53800 (see Fig.~\ref{rxtelc}). 

\begin{figure}
  \includegraphics[angle=0,width=8cm,clip]{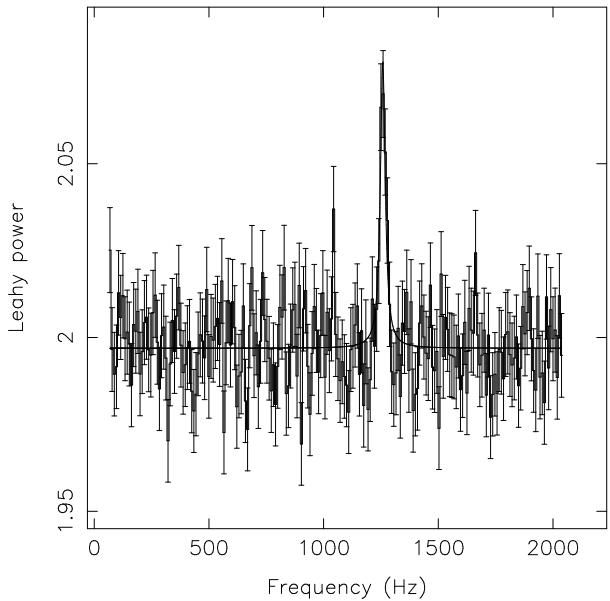}
  
  \caption{ Part of the \citet{ledael1983} normalised average 2--60 keV power spectrum of the three observations indicated
  in Figure~\ref{rxtelc}. The kilohertz QPO is clearly visible. It has a more than 7~$\sigma$ single trial
  significance. The solid line represents the best fitting model consisting of a constant plus a Lorentzian.}

\label{qpo}
\end{figure}

We fitted the 64--2048 Hz part of the average power spectrum of the three observations indicated in Fig.~1 with "kHz QPO"
combined. A fit function consisting of the sum of a constant, to represent the Poisson noise, and a Lorentzian to represent
the kilohertz QPO was used. With a reduced $\chi^2=0.86$ for 207 degrees of freedom the fit was good. The kilohertz QPO has
a frequency of 1258$\pm$2 Hz and a full--width at half maximum of 25$\pm$4 Hz, see Fig.~\ref{qpo} (here and below we report
1~$\sigma$ single parameter errors determined using $\Delta\chi^2=1.0$). The single trial significance in the full 2--60
keV band is 7.5~$\sigma$. Even when taking into account the $\approxlt 1000$ trials we performed, the significance is still
6.8~$\sigma$. The rms amplitude expressed as a fraction of the source count rate in the 2--60 keV band is 17.2$\pm$1.5 per
cent. In order to compare the rms amplitude of the QPO with previous work of e.g.~\citet{2001ApJ...553..335J} we also
calculated power spectra in the 5--60 keV energy band. The fractional rms amplitude in that band is 27$\pm$3 per cent. The
average source count rate in the 2-60 keV band in the three observations containing the kilohertz QPO is 85$\pm$1 counts
per second whereas it is 52$\pm$1 counts per second in the 5--60 keV band. As we will argue in the Discussion the detected
kilohertz QPO can most likely be associated with the upper of the kilohertz QPO pair  found in several LMXBs. In order to
put an upper limit on the presence of a lower kilohertz QPO we determine a 95 per cent confidence upper limit on the
presence of a kilohertz QPO at a frequency 200--400 Hz less than 1258 Hz for a QPO full--width at half maximum of 50, 100,
or 150 Hz. The upper limit on the presence of such a QPO is 20 per cent (2--60 keV). For completeness we did the
same for a kilohertz QPO at a frequency 200-400 Hz higher than 1258 Hz. The upper limit on the presence of such a QPO is 13
per cent (2--60 keV). We also inspected the low--frequency (0.1--256 Hz) part of the average power spectrum containing the
kilohertz QPO. We found a strong (27.4$\pm$1.0 per cent; 2--60 keV) band--limited noise component. We characterised the
band--limited noise component by fitting a Lorentzian to the averaged low--frequency power spectrum. Its central frequency
was 24$\pm$1 Hz, whereas the full--width at half maximum was 24$\pm$2 Hz.

\begin{figure}
  \includegraphics[angle=0,width=8cm,clip]{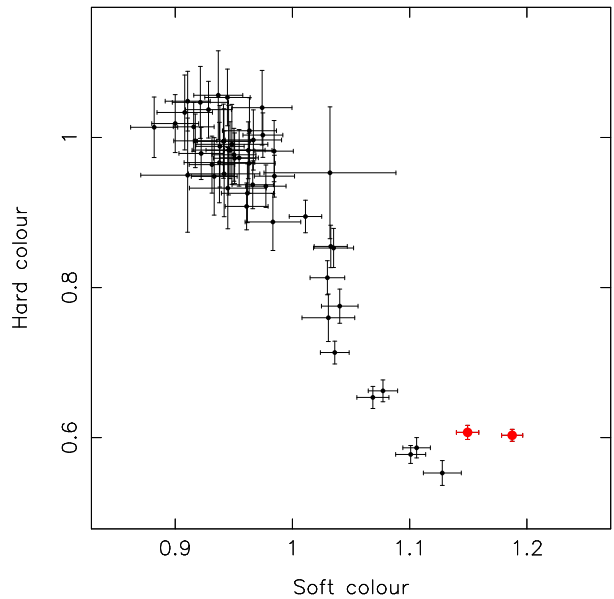}
  
  \caption{ Colour--colour diagram of the Standard 2 PCA data. The small dots are the average colour of the
source during one PCA observation. The large dots at a soft colour between 1.1--1.2 are the
observations when the kilohertz QPO was detected. The soft and hard colour are defined as the ratio between the count
rates in the 3.5--6.4 keV and 2.0--3.5 keV and that between the 9.7--16.0 keV and the 6.4--9.7 keV band,
respectively.}

\label{cd}
\end{figure}

Using the 16~s resolution Standard 2 data we further created colour--colour diagrams (CDs; Figure~\ref{cd}). A soft
colour is plotted vs.~a hard colour; the soft and hard colour are defined as the ratio between the count rates in the
3.5--6.4 keV and 2.0--3.5 keV and that between the 9.7--16.0 keV and the 6.4--9.7 keV band, respectively. The
observed count rates have been corrected for deadtime effects and we have subtracted the background for each
proportional counter unit (PCU) separately. We further corrected the colours for small changes in the instrumental
response (the gain) and for differences in response between the various PCUs, using Crab observations close in time
and assuming the Crab colours to be constant (for a full description of the correction see
\citealt{2003ApJ...596.1155V}). The CD of \src\ bears hallmarks of that of the Island and lower Banana branch of an
'atoll' source (\citealt{hava1989}; see Figure~\ref{cd}). Clearly present is the Island State in the top left part of
Fig.~3 where the source, as in other atoll sources, spends much of its time since motions in the CD are generally
slower here. Due to the low count rate and the fact that often only one or two PCUs were active, we have plotted the
average colour of each observation. Given the short duration of each observation (on average less than 1 ksec), the
chance that the source moved over a large area in the CD is small. The large dots indicate two of the observations
when the kilohertz QPO was present. The elevation above the Earth's limb for the first PCA observation, when the
kilohertz QPO was also present, was too low to create a reliable colour--colour point. The timing properties of that
observation are not affected by the close presence of the Earth's limb.

\section{Discussion}

Using RXTE PCA data we have discovered a kilohertz QPO in the X--ray emission of the ultra--compact X--ray
binary \src. At a frequency of 1258$\pm$2 Hz the kilohertz QPO frequency is rather high in comparison with those
found in other sources (for an overview see \citealt{2006csxs.book...39V}). Furthermore, the fractional rms
amplitude of the kilohertz QPO measured in the 5--60 keV band is the highest found in any kilohertz QPO source.
From the properties of a photospheric radius expansion burst In 't Zand et al.~(in preparation) derive a
distance to the source of $\approx$6 kpc. This implies a persistent source luminosity of $\approx$0.2--0.3 per
cent of the Eddington luminosity, making \src\ one of the weakest LMXBs for which a kilohertz QPO has been
detected; the other similarly weak source is 2S~0918--549 (\citealt{2001ApJ...553..335J}). The high frequency
and high coherence of the kilohertz QPO peak suggests that it can be identified as the upper kilohertz QPO. The
lower kilohertz QPO has never been detected at frequencies higher than $\approx$1000 Hz when both kilohertz QPO
peaks were detected simultaneously (\citealt{2006csxs.book...39V}) and the coherence of the upper kilohertz QPO
is found to increase at frequencies above $\approx$1100 Hz (\citealt{2006MNRAS.371.1925M,2006MNRAS.370.1140B}).
The spectral properties suggest that \src\ belongs to the class of 'atoll' sources. The kilohertz QPO was only
significantly detected when the source was in the lower banana branch similar to what has been found in other
atoll sources (\citealt{mevafo1999}; \citealt{vafova2000}).

The high frequency yields the exciting possibility to obtain constraints on the neutron star. For instance, if a
second simultaneous kilohertz QPO is discovered, the neutron star spin rate can be constrained.  In the 8 sources
where both twin kilohertz QPOs and neutron star spins were directly measured, $\Delta\nu\approx\nu_{\rm spin}$ or
$\Delta\nu\approx\nu_{\rm spin}/2$ (see review by \citealt{2006csxs.book...39V}). Another constraint involves the
neutron star mass and radius. The measured 1258 Hz frequency is high. Only in one object has a higher frequency ever
been measured (see Section 1). If one interprets the observed QPO frequency as a Keplerian frequency and requires
that the neutron star is smaller than the radius associated with that Keplerian frequency a neutron star
mass--dependent limit can be put on the neutron star radius. In addition if one assumes that the innermost stable
circular orbit has a radius also smaller than the associated Keplerian radius a limit on the neutron star mass is
obtained (\citealt{milaps1998}). Assuming a non--rotating neutron star this yields the following limits on the mass
and radius of the neutron star for the 1258 Hz QPO: M${\rm _{NS}}<$1.75 M$_\odot$ and R$<$15.5 km.

As mentioned above the fractional rms amplitude of the kilohertz QPO is the highest found to date. Similarly, the
fractional rms amplitude of the broad low--frequency noise component is very high at 27 per cent (2--60 keV). This all
fits--in with the trend of increasing rms amplitude for a decreasing source luminosity as shown for the upper kilohertz
QPO before by \citet{2001ApJ...553..335J}. Using the data from \citet{2001ApJ...553..335J} and the data on \src\ we
have fitted the relation between the fractional rms amplitude of the upper kilohertz QPO and source luminosity. A
power--law fit gives an index of $-0.35\pm$0.01. From figure 3 in \citet{2006MNRAS.371.1925M} it seems as if the
increase in the maximum fractional rms amplitude of a source with decreasing source luminosity levels off at 18--20 per
cent (measured in the 2--60 keV band). However, the kilohertz QPO that we found in \src\ does not seem to be compatible
with this trend. Even though the measured amplitude in the 2--60 keV band of 17.2$\pm$1.5 per cent is close to this
saturation level, this amplitude is measured while the QPO frequency was very high (1258 Hz). In other sources the rms
amplitude of the upper kilohertz QPO decreases strongly at upper kilohertz QPO frequencies above 700--800 Hz. Hence,
either the relation between kilohertz QPO frequency and fractional rms amplitude is significantly different in \src\
from that in other sources or the increase in fractional rms amplitude and source luminosity does not level off at
18--20 per cent but keeps increasing.

\section*{Acknowledgments}  

\noindent PGJ acknowledges support from the Netherlands Organisation for Scientific Research and R.~Cornelisse for
comments on an earlier version of the Manuscript.


\begin{thebibliography}{27}
\expandafter\ifx\csname natexlab\endcsname\relax\def\natexlab#1{#1}\fi

\bibitem[{{Arnaud}(1996)}]{ar1996}
{Arnaud}, K.~A., 1996, in ASP Conf. Ser. 101: Astronomical Data Analysis
  Software and Systems V, vol.~5, p.~17

\bibitem[{{Barret} et~al.(2006){Barret}, {Olive}, \&
  {Miller}}]{2006MNRAS.370.1140B}
{Barret}, D., {Olive}, J.-F., {Miller}, M.~C., 2006, \mnras, 370, 1140

\bibitem[{{Bassa} et~al.(2006){Bassa}, {Jonker}, {in't Zand}, \&
  {Verbunt}}]{2006A&A...446L..17B}
{Bassa}, C.~G., {Jonker}, P.~G., {in't Zand}, J.~J.~M., {Verbunt}, F., 2006,
  \aap, 446, L17

\bibitem[{{Boller} et~al.(1997){Boller}, {Haberl}, {Voges}, {Piro}, \&
  {Heise}}]{1997IAUC.6546....1B}
{Boller}, T., {Haberl}, F., {Voges}, W., {Piro}, L., {Heise}, J., 1997,
  \iaucirc, 6546, 1

\bibitem[{{Carpenter} et~al.(1977){Carpenter}, {Eyles}, {Skinner}, {Wilson}, \&
  {Willmore}}]{1977MNRAS.179P..27C}
{Carpenter}, G.~F., {Eyles}, C.~J., {Skinner}, G.~K., {Wilson}, A.~M.,
  {Willmore}, A.~P., 1977, \mnras, 179, 27P

\bibitem[{{Forman} et~al.(1978){Forman}, {Jones}, {Cominsky}, {Julien},
  {Murray}, {Peters}, {Tananbaum}, \& {Giacconi}}]{1978ApJS...38..357F}
{Forman}, W., {Jones}, C., {Cominsky}, L., {Julien}, P., {Murray}, S.,
  {Peters}, G., {Tananbaum}, H., {Giacconi}, R., 1978, \apjs, 38, 357

\bibitem[{{Hasinger} \& {van der Klis}(1989)}]{hava1989}
{Hasinger}, G., {van der Klis}, M., 1989, \aap, 225, 79

\bibitem[{{in 't Zand} et~al.(2007){in 't Zand}, {Jonker}, \&
  {Markwardt}}]{2007astro.ph..1810I}
{in 't Zand}, J.~J.~M., {Jonker}, P.~G., {Markwardt}, C.~B., 2007, ArXiv
  Astrophysics e-prints

\bibitem[{{Jonker} et~al.(2001)}]{2001ApJ...553..335J}
{Jonker}, P.~G., et~al., 2001, \apj, 553, 335

\bibitem[{{Kong}(2006)}]{2006ATel..875....1K}
{Kong}, A.~K.~H., 2006, The Astronomer's Telegram, 875, 1

\bibitem[{{Leahy} et~al.(1983){Leahy}, {Darbro}, {Elsner}, {Weisskopf}, {Kahn},
  {Sutherland}, \& {Grindlay}}]{ledael1983}
{Leahy}, D.~A., {Darbro}, W., {Elsner}, R.~F., {Weisskopf}, M.~C., {Kahn}, S.,
  {Sutherland}, P.~G., {Grindlay}, J.~E., 1983, \apj, 266, 160

\bibitem[{{M{\'e}ndez}(2006)}]{2006MNRAS.371.1925M}
{M{\'e}ndez}, M., 2006, \mnras, 371, 1925

\bibitem[{{M{\'e}ndez} et~al.(1999){M{\'e}ndez}, {van der Klis}, {Ford},
  {Wijnands}, \& {van Paradijs}}]{mevafo1999}
{M{\'e}ndez}, M., {van der Klis}, M., {Ford}, E.~C., {Wijnands}, R., {van
  Paradijs}, J., 1999, \apjl, 511, L49

\bibitem[{{Miller} et~al.(1998){Miller}, {Lamb}, \& {Psaltis}}]{milaps1998}
{Miller}, M.~C., {Lamb}, F.~K., {Psaltis}, D., 1998, \apj, 508, 791

\bibitem[{{Nelemans} \& {Jonker}(2006)}]{2006astro.ph..5722N}
{Nelemans}, G., {Jonker}, P.~G., 2006, ArXiv Astrophysics e-prints

\bibitem[{{Nelson} et~al.(1986){Nelson}, {Rappaport}, \&
  {Joss}}]{1986ApJ...304..231N}
{Nelson}, L.~A., {Rappaport}, S.~A., {Joss}, P.~C., 1986, \apj, 304, 231

\bibitem[{{Piro} et~al.(1997)}]{1997IAUC.6538....2P}
{Piro}, L., et~al., 1997, \iaucirc, 6538, 2

\bibitem[{{Romano} et~al.(2006)}]{2006GCN..5436....1R}
{Romano}, P., et~al., 2006, GRB Coordinates Network, 5436, 1

\bibitem[{{Savonije} et~al.(1986){Savonije}, {de Kool}, \& {van den
  Heuvel}}]{1986AA...155...51S}
{Savonije}, G.~J., {de Kool}, M., {van den Heuvel}, E.~P.~J., 1986, \aap, 155,
  51

\bibitem[{{Seward} et~al.(1976){Seward}, {Page}, {Turner}, \&
  {Pounds}}]{1976MNRAS.177P..13S}
{Seward}, F.~D., {Page}, C.~G., {Turner}, M.~J.~L., {Pounds}, K.~A., 1976,
  \mnras, 177, 13P

\bibitem[{{Strohmayer} \& {Bildsten}(2006)}]{2006csxs.book..113S}
{Strohmayer}, T., {Bildsten}, L., 2006, {New views of thermonuclear bursts},
  Compact stellar X-ray sources, p. 113

\bibitem[{{van der Klis}(2006{\natexlab{a}})}]{2006AdSpR..38.2675V}
{van der Klis}, M., 2006{\natexlab{a}}, Advances in Space Research, 38, 2675

\bibitem[{{van der Klis}(2006{\natexlab{b}})}]{2006csxs.book...39V}
{van der Klis}, M., 2006{\natexlab{b}}, {Rapid X-ray Variability}, Compact
  stellar X-ray sources, p.~39

\bibitem[{{van Straaten} et~al.(2000){van Straaten}, {Ford}, {van der Klis},
  {M{\'e}ndez}, \& {Kaaret}}]{vafova2000}
{van Straaten}, S., {Ford}, E.~C., {van der Klis}, M., {M{\'e}ndez}, M.,
  {Kaaret}, P., 2000, \apj, 540, 1049

\bibitem[{{van Straaten} et~al.(2002){van Straaten}, {van der Klis}, {di
  Salvo}, \& {Belloni}}]{2002ApJ...568..912V}
{van Straaten}, S., {van der Klis}, M., {di Salvo}, T., {Belloni}, T., 2002,
  \apj, 568, 912

\bibitem[{{van Straaten} et~al.(2003){van Straaten}, {van der Klis}, \&
  {M{\'e}ndez}}]{2003ApJ...596.1155V}
{van Straaten}, S., {van der Klis}, M., {M{\'e}ndez}, M., 2003, \apj, 596, 1155

\bibitem[{{Wijnands} et~al.(2003){Wijnands}, {van der Klis}, {Homan},
  {Chakrabarty}, {Markwardt}, \& {Morgan}}]{2003Natur.424...44W}
{Wijnands}, R., {van der Klis}, M., {Homan}, J., {Chakrabarty}, D.,
  {Markwardt}, C.~B., {Morgan}, E.~H., 2003, \nat, 424, 44

\end{thebibliography}
\end{document}